\documentclass[conference]{IEEEtran}
\IEEEoverridecommandlockouts
\usepackage{cite}
\usepackage{amsmath,amssymb,amsfonts}
\usepackage[nolist]{acronym}
\usepackage{algorithmic}
\usepackage{hyperref}
\usepackage{scalerel}
\usepackage{graphicx}
\usepackage{textcomp}
\usepackage{xcolor}
\usepackage{tikz}
\usepackage{nicefrac}

\usetikzlibrary{calc}
\usepackage{pgfplots}
\usepgfplotslibrary{groupplots}
\usepgfplotslibrary{colorbrewer}

\makeatletter
\let\MYcaption\@makecaption
\makeatother

\usepackage[font=footnotesize]{subcaption}

\makeatletter
\let\@makecaption\MYcaption
\makeatother

\DeclareMathOperator*{\argmax}{arg\,max}
\DeclareMathOperator*{\argmin}{arg\,min}

\usetikzlibrary{svg.path}

\def\BibTeX{{\rm B\kern-.05em{\sc i\kern-.025em b}\kern-.08em
    T\kern-.1667em\lower.7ex\hbox{E}\kern-.125emX}}

\definecolor{orcidlogocol}{HTML}{A6CE39}
\tikzset{
  orcidlogo/.pic={
    \fill[orcidlogocol] svg{M256,128c0,70.7-57.3,128-128,128C57.3,256,0,198.7,0,128C0,57.3,57.3,0,128,0C198.7,0,256,57.3,256,128z};
    \fill[white] svg{M86.3,186.2H70.9V79.1h15.4v48.4V186.2z}
                 svg{M108.9,79.1h41.6c39.6,0,57,28.3,57,53.6c0,27.5-21.5,53.6-56.8,53.6h-41.8V79.1z M124.3,172.4h24.5c34.9,0,42.9-26.5,42.9-39.7c0-21.5-13.7-39.7-43.7-39.7h-23.7V172.4z}
                 svg{M88.7,56.8c0,5.5-4.5,10.1-10.1,10.1c-5.6,0-10.1-4.6-10.1-10.1c0-5.6,4.5-10.1,10.1-10.1C84.2,46.7,88.7,51.3,88.7,56.8z};
  }
}

\newcommand\orcidicon[1]{\href{https://orcid.org/#1}{\mbox{\scalerel*{
\begin{tikzpicture}[yscale=-1,transform shape]
\pic{orcidlogo};
\end{tikzpicture}
}{|}}}}

\begin{acronym}
 \acro{R2M}{raw 2\textsuperscript{nd} moment}
 \acro{CSI}{channel state information}
 \acro{UE}{user equipment}
 \acro{EM}{electromagnetic}
 \acro{UL}{uplink}
 \acro{BS}{base station}
 \acro{TDD}{time division duplex}
 \acro{FDD}{frequency division duplex}
 \acro{ECC}{error-correcting code}
 \acro{MLD}{maximum likelihood decoding}
 \acro{HDD}{hard decision decoding}
 \acro{IF}{intermediate frequency}
 \acro{RF}{radio frequency}
 \acro{SDD}{soft decision decoding}
 \acro{NND}{neural network decoding}
 \acro{CNN}{convolutional neural network}
 \acro{ML}{maximum likelihood}
 \acro{GPU}{graphical processing unit}
 \acro{BP}{belief propagation}
 \acro{LTE}{Long Term Evolution}
 \acro{BER}{bit error rate}
 \acro{SNR}{signal-to-noise-ratio}
 \acro{ReLU}{rectified linear unit}
 \acro{BPSK}{binary phase shift keying}
 \acro{QPSK}{quadrature phase shift keying}
 \acro{AWGN}{additive white Gaussian noise}
 \acro{MSE}{mean squared error}
 \acro{LLR}{log-likelihood ratio}
 \acro{MAP}{maximum a posteriori}
 \acro{NVE}{normalized validation error}
 \acro{BCE}{binary cross-entropy}
 \acro{CE}{cross-entropy}
 \acro{BLER}{block error rate}
 \acro{SQR}{signal-to-quantisation-noise-ratio}
 \acro{MIMO}{multiple-input multiple-output}
 \acro{mMIMO}{massive multiple-input multiple-output}
 \acro{OFDM}{orthogonal frequency division multiplex}
 \acro{RF}{radio frequency}
 \acro{LoS}{line of sight}
 \acro{NLoS}{non-line of sight}
 \acro{NMSE}{normalized mean squared error}
 \acro{CFO}{carrier frequency offset}
 \acro{SFO}{sampling frequency offset}
 \acro{IPS}{indoor positioning system}
 \acro{TRIPS}{time-reversal IPS}
 \acro{RSSI}{received signal strength indicator}
 \acro{MIMO}{multiple-input multiple-output}
 \acro{ENoB}{effective number of bits}
 \acro{AGC}{automatic gain control}
 \acro{ADC}{analog to digital converter}
 \acro{ADCs}{analog to digital converters}
 \acro{FB}{front bandpass}
 \acro{FPGA}{field programmable gate array}
 \acro{JSDM}{Joint Spatial Division and Multiplexing}
 \acro{NN}{neural network}
 \acro{IF}{intermediate frequency}
 \acro{LoS}{line-of-sight}
 \acro{NLoS}{non-line-of-sight}
 \acro{DSP}{digital signal processing}
 \acro{AFE}{analog front end}
 \acro{SQNR}{signal-to-quantisation-noise-ratio}
 \acro{SINR}{signal-to-interference-noise-ratio}
 \acro{ENoB}{effective number of bits}
 \acro{PCB}{printed circuit board}
 \acro{EVM}{error vector mangnitude}
 \acro{CDF}{cumulative distribution function}
 \acro{MRC}{maximum ratio combining}
 \acro{MRP}{maximum ratio precoding}
 \acro{MRT}{maximum ratio transmission}
 \acro{DeepL}{deep-learning}
 \acro{DL}{downlink}
 \acro{SISO}{single-input single-output}
 \acro{SGD}{stochastic gradient descent}
 \acro{CP}{cyclic prefix}
 \acro{MISO}{Multiple Input Single Output}
 \acro{LMMSE}{linear minimum mean square error}
 \acro{ZF}{zero forcing}
 \acro{USRP}{universal software radio peripheral}
 \acro{RNN}{recurrent neural network}
 \acro{GRU}{gated recurrent unit}
 \acro{LSTM}{long short-term memory}
 \acro{NTM}{neural turing machine}
 \acro{DNC}{differentiable neural computer}
 \acro{TCN}{temporal convolutional network}
 \acro{FCL}{fully connected layer}
 \acro{MANN}{memory augmented neural network}
 \acro{RNN}{recurrent neural network}
 \acro{DNN}{deep neural network}
 \acro{FIR}{finite impulse response}
 \acro{BPTT}{back-propagation through time}
 \acro{GAN}{generative adversarial network}
 \acro{ELU}{exponential linear unit}
 \acro{tanh}{hyperbolic tangent}
 \acro{BICM}{bit-interleaved coded modulation}
 \acro{OTA}{over-the-air}
 \acro{IM}{intensity modulation}
 \acro{DD}{direct detection}
 \acro{RL}{reinforcement learning}
 \acro{SDR}{software-defined radio}
 \acro{WGAN}{Wasserstein generative adversarial network}
 \acro{BMD}{bit-metric decoding}
 \acro{BMI}{bit-wise mutual information}
 \acro{LDPC}{low-density parity-check}
 \acro{IDD}{iterative demapping and decoding}
 \acro{JSD}{Jensen-Shannon divergence}
 \acro{MMSE}{minimum mean square error}
 \acro{FFT}{fast Fourier transform}
 \acro{DFT}{discrete Fourier transform}
 \acro{IFFT}{inverse fast Fourier transform}
 \acro{QAM}{quadrature amplitude modulation}
 \acro{EMD}{earth mover's distance}
 \acro{TDL}{tapped delay line}
 \acro{KL}{Kullback-Leibler}
 \acro{PRACH}{physical random access channel}
 \acro{URLLC}{ultra-reliable low-latency communication}
 \acro{ANOMA}{asynchronous non-orthogonal multiple access}
 \acro{FEC}{forward error correction}
 \acro{PAPR}{peak-to-average power ratio}
 \acro{APP}{a posteriori probability}
 \acro{COTS}{commercial off-the-shelf}
 \acro{PLL}{phase locked loop}
 \acro{STO}{sampling time offset}
 \acro{SFO}{sampling frequency offset}
 \acro{CFO}{carrier frequency offset}
 \acro{CPO}{carrier phase offset}
 \acro{CSI}{channel state information}
 \acro{GNSS}{global navigation satellite system}
 \acro{ELAA}{extremely large aperture array}
 \acro{UE}{user equipment}
 \acro{DICHASUS}{\underline{Di}stributed \underline{Cha}nnel \underline{S}ounder by \underline{U}niversity of \underline{S}tuttgart}
 \acro{JCaS}{Joint Communication and Sensing}
 \acro{CT}{continuity}
 \acro{TW}{trustworthiness}
 \acro{KS}{Kruskal's stress}
 \acro{PCA}{principal component analysis}
 \acro{AoA}{angle of arrival}
 \acro{ToA}{time of arrival}
 \acro{TDoA}{time difference of arrival}
 \acro{ToT}{time of transmission}
 \acro{RD}{Rajski's distance}
 \acro{ADP}{angle-delay profile}
 \acro{MPC}{multipath component}
 \acro{CIR}{channel impulse response}
 \acro{MAE}{mean absolute error}
 \acro{MDS}{multidimensional scaling}
 \acro{t-SNE}{t-distributed stochastic neighbor embedding}
 \acro{SM}{Sammon's mapping}
 \acro{CIRA}{channel impulse response amplitude}
 \acro{CS}{cosine similarity}
 \acro{FCF}{forward charting function}
 \acro{CMD}{correlation matrix distance}
 \acro{UWB}{ultra-wideband}
 \acro{MUSIC}{MUltiple SIgnal Classification}
 \acro{FBCM}{forward-backward correlation matrix}
 \acro{MDL}{minimum descriptive length}
 \acro{BFGS}{Broyden-Fletcher-Goldfarb-Shanno}
 \acro{UPA}{uniform planar array}
 \acro{RMS}{root mean square}
 \acro{DRMS}{distance root mean square}
 \acro{CEP}{circular error probable}
 \acro{R95}{95\textsuperscript{th} percentile radius}
 \acro{eCDF}{empirical cumulative distribution function}
\end{acronym}

\definecolor{mittelblau}{RGB}{0, 126, 198}
\definecolor{violettblau}{cmyk}{0.9, 0.6, 0, 0}
\definecolor{rot}{RGB}{238, 28 35}
\definecolor{apfelgruen}{RGB}{140, 198, 62}
\definecolor{gelb}{RGB}{1, 221, 0}
\definecolor{orange}{RGB}{244, 111, 33}
\definecolor{pink}{RGB}{237, 0, 140}
\definecolor{lila}{RGB}{128, 10, 145}
\definecolor{hellgrau}{RGB}{224, 224, 224}
\definecolor{mittelgrau}{RGB}{128, 128, 128}
\definecolor{dunkelgrau}{RGB}{80,80,80}
\definecolor{anthrazit}{RGB}{19, 31, 31}

\begin{document}

\title{Augmenting Channel Charting with Classical Wireless Source Localization Techniques
\thanks{This work is supported by the German Federal Ministry of Education and Research (BMBF) within the projects Open6GHub (grant no. 16KISK019) and KOMSENS-6G (grant no. 16KISK113).}}

\author{\IEEEauthorblockN{Florian Euchner\textsuperscript{\orcidicon{0000-0002-8090-1188}}, Phillip Stephan\textsuperscript{\orcidicon{0009-0007-4036-668X}}, Stephan ten Brink\textsuperscript{\orcidicon{0000-0003-1502-2571}} \\}

\IEEEauthorblockA{
Institute of Telecommunications, Pfaffenwaldring 47, University of  Stuttgart, 70569 Stuttgart, Germany \\ \{euchner,stephan,tenbrink\}@inue.uni-stuttgart.de
}
}

\maketitle

\begin{abstract}
Channel Charting aims to construct a map of the radio environment by leveraging similarity relationships found in high-dimensional channel state information.
Although resulting channel charts usually accurately represent local neighborhood relationships, even under conditions with strong multipath propagation, they often fall short in capturing global geometric features.
On the other hand, classical model-based localization methods, such as triangulation and multilateration, can easily localize signal sources in the global coordinate frame.
However, these methods rely heavily on the assumption of line-of-sight channels and distributed antenna deployments.
Based on measured data, we compare classical source localization techniques to channel charts with respect to localization performance.
We suggest and evaluate methods to enhance Channel Charting with model-based localization approaches:
One approach involves using information derived from classical localization methods to map channel chart locations to physical positions after conventional training of the forward charting function.
Foremost, though, we suggest to incorporate information from model-based approaches during the training of the forward charting function in what we call ``augmented Channel Charting''.
We demonstrate that Channel Charting can outperform classical localization methods on the considered dataset.
\end{abstract}

\begin{IEEEkeywords}
Channel Charting, massive MIMO, localization
\end{IEEEkeywords}

\section{Introduction}
Wireless localization systems are commonplace:
Smartphones and cars navigate thanks to \acp{GNSS}, lost items can be found with the help of \ac{UWB} tags, Bluetooth beacons enable localization of assets in factories and emergency responders can access approximate caller locations determined by the cellular network.
Many common wireless localization systems rely on a few common of techniques to determine the position of \acp{UE} (like smartphones, becons or cars), namely \ac{AoA} estimation and subsequent triangulation, \ac{ToA} estimation and subsequent multilateration, receive power-based multilateration or a combination of aforementioned methods.
What these localization techniques have in common is that they are \emph{model-based}:
They rely on the assumption of a certain channel model, for example relying on the existence of a sufficiently strong \ac{LoS} channel, or assuming a certain fading model with a specific path loss coefficient.

More recently, the concept of Channel Charting has been introduced \cite{studer_cc} as a data-driven approach that learns a so-called channel chart, a kind of map of the radio environment.
One of the initially proposed applications of Channel Charting was \emph{relative} localization, but additional work suggested various methods to also enable localization in the absolute global coordinate frame \cite{taner2023channel} \cite{pihlajasalo2020absolute}.
In contrast to model-based techniques, Channel Charting relies on similarity relationships between measured \ac{CSI} and thereby specifically exploits the multipath characteristics of the propagation environment.

So far, Channel Charting and model-based localization have only been considered separately, but were never compared to each other or even combined.
In this work, we address this gap by directly comparing and combining model-based approaches and Channel Charting.
All evaluations are performed on one particular dataset of \ac{CSI} measurements that is introduced in Section \ref{sec:dataset}.
In Section \ref{sec:classical}, we explain the classical model-based localization methods and evaluate their performance, and in Section \ref{sec:cc}, we evaluate the performance of Channel Charting-based localization.
Subsequently, in Section \ref{sec:augmented}, we combine classical techniques with Channel Charting into what we call \emph{augmented} Channel Charting, and finally, in Section \ref{sec:comparison}, we compare the performance of all methods with respect to various metrics.
The datasets and source code for both classical and Channel Charting-based localization used by this work are publicly available\footnote{\url{https://github.com/Jeija/ToA-AoA-Augmented-ChannelCharting/}}.

\section{Dataset and System Model}
\label{sec:dataset}

\begin{figure*}
    \centering
    \begin{subfigure}[b]{0.33\textwidth}
        \centering
        \includegraphics[width=0.95\textwidth, trim = 30 200 30 0, clip]{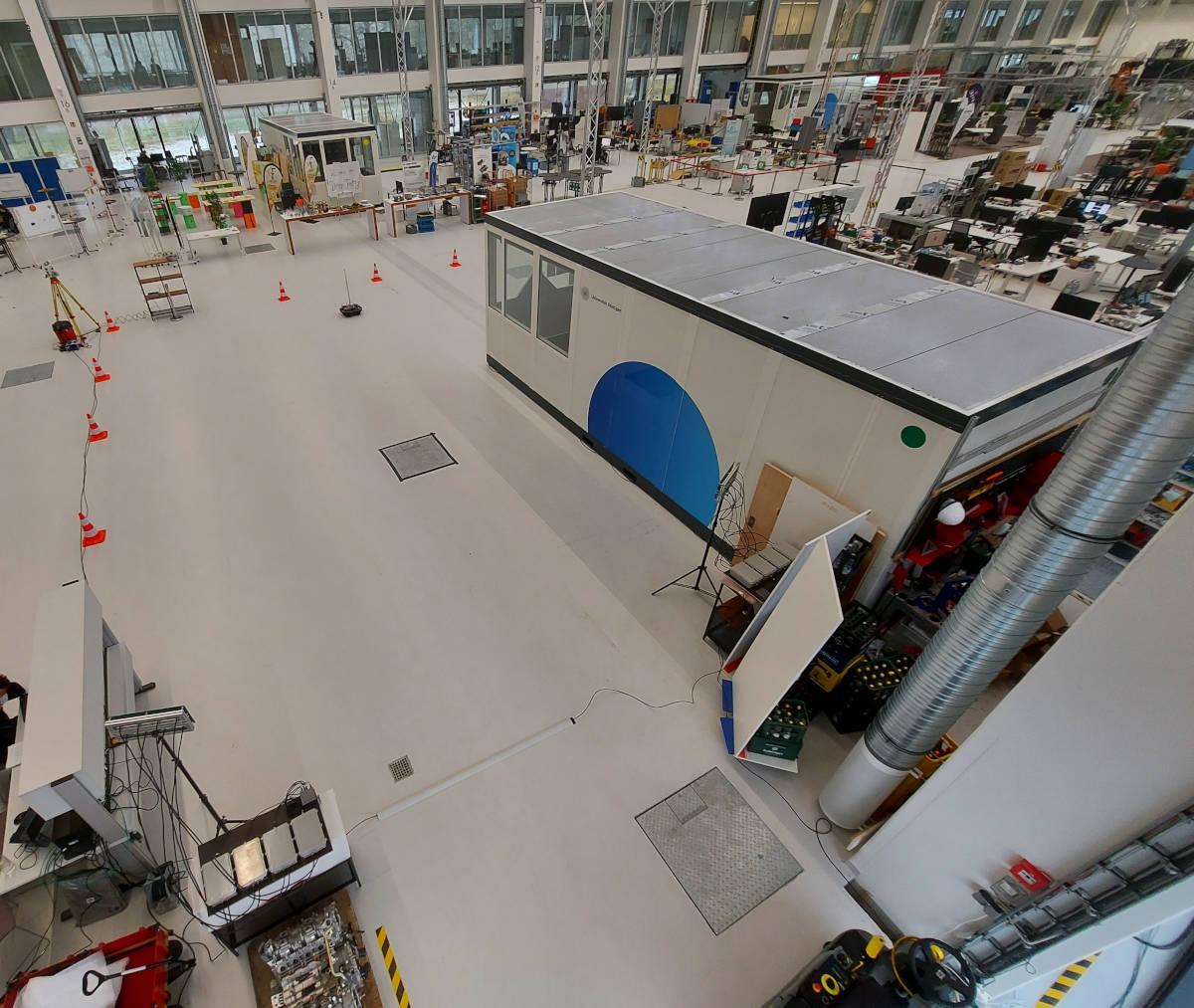}
        \vspace{0.45cm}
        \caption{}
    \end{subfigure}
    \begin{subfigure}[b]{0.32\textwidth}
        \centering
        \begin{tikzpicture}
            \begin{axis}[
                width=0.729\columnwidth,
                height=0.6\columnwidth,
                scale only axis,
                xmin=-15.5,
                xmax=6.1,
                ymin=-18.06,
                ymax=-1.5,
                xlabel = {Coordinate $x_1 ~ [\mathrm{m}]$},
                ylabel = {Coordinate $x_2 ~ [\mathrm{m}]$},
                ylabel shift = -8 pt,
                xlabel shift = -4 pt,
                xtick={-10, -6, -2, 2}
            ]
                \addplot[thick,blue] graphics[xmin=-14.5,ymin=-17.06,xmax=4.1,ymax=-1.5] {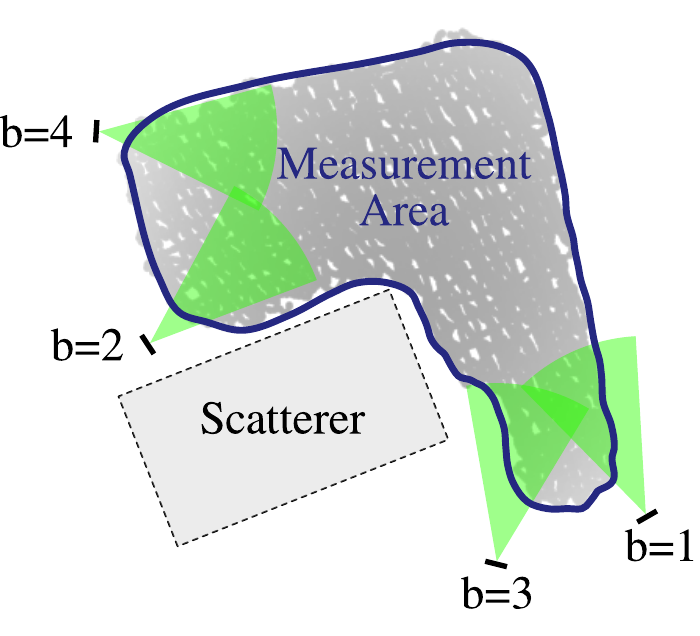};
            \end{axis}
        \end{tikzpicture}
        \vspace{-0.5cm}
        \caption{}
        \label{fig:labelled-area}
    \end{subfigure}
    \begin{subfigure}[b]{0.32\textwidth}
        \centering
        \begin{tikzpicture}
            \begin{axis}[
                width=0.6\columnwidth,
                height=0.6\columnwidth,
                scale only axis,
                xmin=-12.5,
                xmax=2.5,
                ymin=-14.5,
                ymax=-1.5,
                xlabel = {Coordinate $x_1 ~ [\mathrm{m}]$},
                ylabel = {Coordinate $x_2 ~ [\mathrm{m}]$},
                ylabel shift = -8 pt,
                xlabel shift = -4 pt,
                xtick={-10, -6, -2, 2}
            ]
                \addplot[thick,blue] graphics[xmin=-12.5,ymin=-14.5,xmax=2.5,ymax=-1.5] {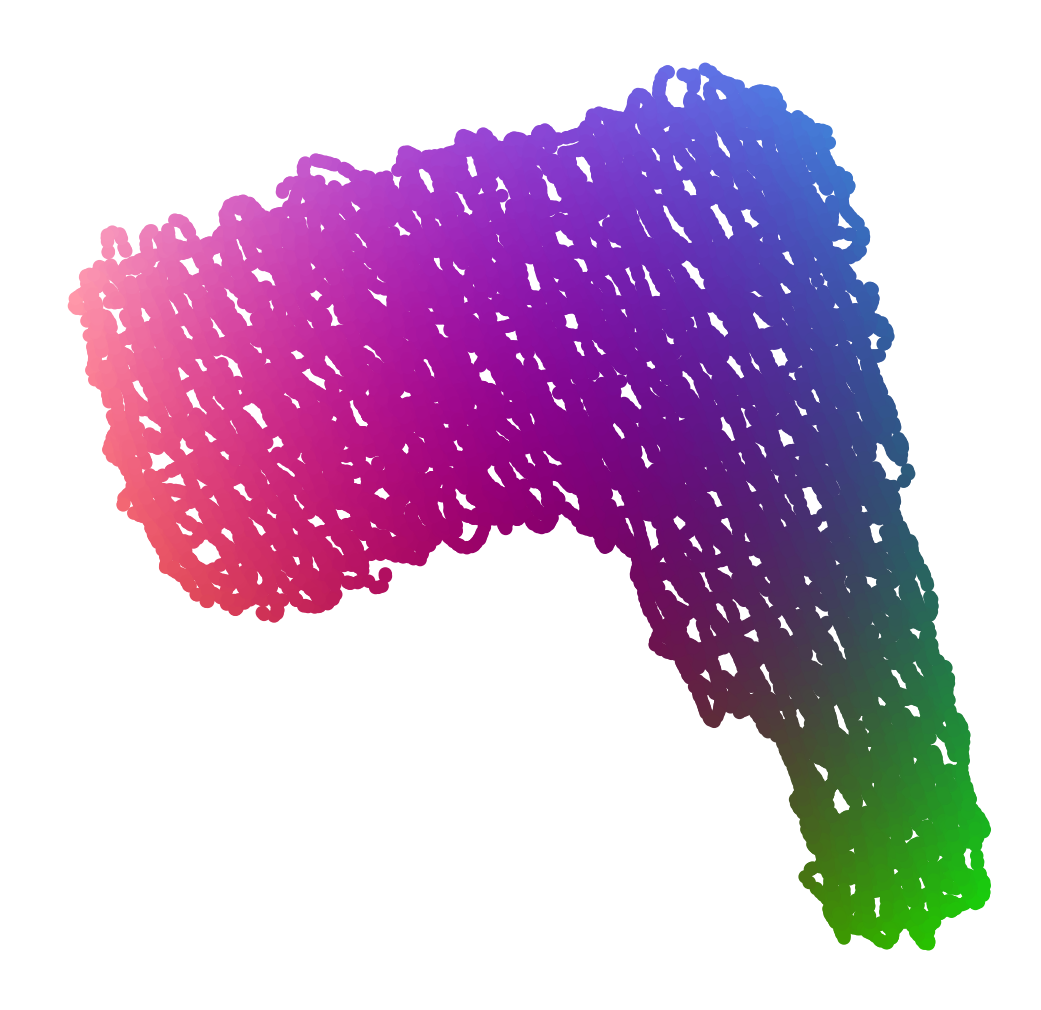};
            \end{axis}
        \end{tikzpicture}
        \vspace{-0.1cm}
        \caption{}
        \label{fig:groundtruth-map}
    \end{subfigure}
    \vspace{-0.3cm}
    \caption{Information about the environment the dataset was measured in: The figure shows (a) a photograph of the environment, (b) a top view map and (c) a scatter plot of colorized ``ground truth'' positions of datapoints in $\mathcal S_\mathrm{full}$, colorized with the measured delay spread. The antenna arrays in the map are drawn to scale as black rectangles and their viewing direction is indicated by the green sectors.}
    \label{fig:industrial_environment}
\end{figure*}

As in our previous work on Channel Charting \cite{stephan2023angle}, we learn the channel chart for a dataset generated by \emph{\ac{DICHASUS}}, our distributed \ac{mMIMO} channel sounder, whose architecture is thoroughly described in \cite{dichasus2021}.
In brief, \ac{DICHASUS} measures the propagation channel between a single transmitter and many receive antennas.
It achieves long-term phase-coherence, even if antennas are distributed over a wide area.
\ac{DICHASUS} provides large datasets containing frequency-domain \ac{CSI} tensors $\mathbf H^{(l)}$, alongside side information like timestamps and accurate information about the positions of all antennas.

The dataset chosen for the following analyses is called \emph{dichasus-cf0x} \cite{dataset-dichasus-cf0x}, and was captured in an industrial environment with $B = 4$ separate \acp{UPA} with half-wavelength antenna spacing, made up of $M_\mathrm{row} \times M_\mathrm{col} = 2 \times 4$ antennas each.
$N_\mathrm{sub} = 1024$ \ac{OFDM} channel coefficients were measured at a carrier frequency of $1.272\,\mathrm{GHz}$ and with a total bandwidth of $50\,\mathrm{MHz}$.
The complex-valued channel coefficients of all arrays, antennas and subcarriers for time instance $l$ are stored in the tensor $\mathbf H^{(l)} \in \mathbb C^{B \times M_\mathrm{row} \times M_\mathrm{col} \times N_\mathrm{sub}}$.
For example, $H_{2,1,3,400}$ is the subcarrier coefficient of the 400\textsuperscript{th} subcarrier of the third antenna in the first row of antenna array $b = 2$.

The single dipole transmit antenna is mounted on top of a robot, which travels along a set of trajectories inside a defined, L-shaped area, with an overall bounding box size of approximately $14\,\mathrm{m} \times 14\,\mathrm{m}$.
A prism is attached to the tip of the antenna and tracked with millimeter-level precision by a tachymeter, providing ``ground truth'' positions $\mathbf x^{(l)}$.
In addition to \ac{CSI} and reference positions, timestamps $t^{(l)}$ are recorded.
Thus, the complete dataset can be represented as the following set containing a total of $L$ datapoints:
\[
    \text{Dataset}: \mathcal S = \left\{ \left(\mathbf H^{(l)}, \mathbf x^{(l)}, t^{(l)} \right) \right\}_{l = 1, \ldots, L}
\]

A photo and a top view map of the environment are shown in Fig. \ref{fig:industrial_environment}.
A large metal container is located at the inner corner of the L-shape, blocking the \ac{LoS}.
The true datapoint positions $\mathbf x^{(l)}$ are shown in Fig. \ref{fig:industrial_environment}.
The points have been colorized and the datapoints will retain their color even as the FCF maps them to a position in the channel chart.
This allows for a visual evaluation of the generated chart: If the global topology is preserved, a similar color gradient should appear in the chart.
If the local geometry is preserved, only datapoints with similar colors should be close to each other in the channel chart.

Thanks to synchronization and the distributed setup of the antenna arrays, time of arrival-based multilateration should be possible in principle (though limited in accuracy by the bandwidth of the system).
Also, since there are $M_\mathrm{row} = 4$ antennas in one row of an antenna array, the azimuth angle of arrival can be estimated and triangulation should be possible.
The centers of the $B = 4$ antenna arrays are known to be at $\mathbf z^{(b)} \in \mathbb R^3$, and the antenna normal vectors are $\mathbf n^{(b)} \in \mathbb R^3$.

In all localization tasks, we assume that the height of the transmitter is known and constant, i.e., we provide $x_3^{(l)}$ to the estimation algorithm.
This simplifies the estimation task to two-dimensional localization.
This assumption of a constant transmitter height is approximately true, because the robot always remains on the factory floor and that floor is almost perfectly flat.

\section{Classical Localization Techniques}
\label{sec:classical}

\begin{figure*}
    \centering
    \begin{subfigure}{0.3\textwidth}
        \centering
        \begin{tikzpicture}
            \begin{axis}[
                width=0.6\columnwidth,
                height=0.6\columnwidth,
                scale only axis,
                enlargelimits=false,
                axis on top,
                xlabel = {Coordinate $\hat x_1 ~ [\mathrm{m}]$},
                ylabel = {Coordinate $\hat x_2 ~ [\mathrm{m}]$},
                ylabel shift = -8 pt,
                xlabel shift = -4 pt,
                xtick={-15, -10, -5, 0, 5}
            ]
                \addplot graphics[xmin=-16.84,xmax=6.60,ymin=-18.53,ymax=2.96] {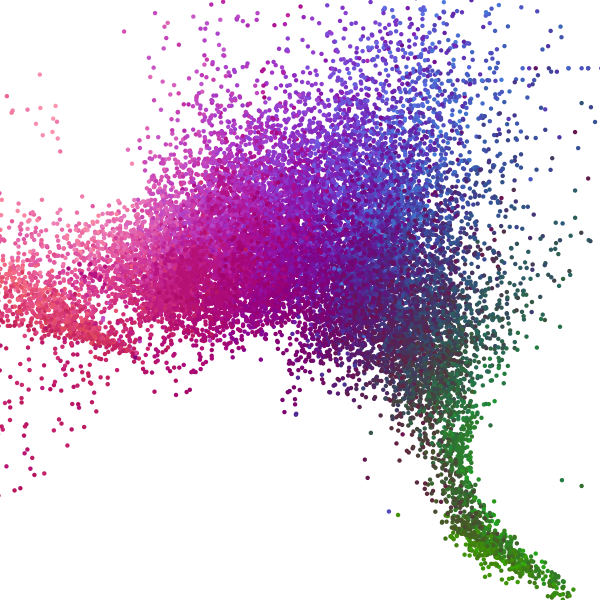};
            \end{axis}
        \end{tikzpicture}
        \vspace{-0.1cm}
        \caption{ToA / Multilateration: $\mathcal L^{(l)}_\mathrm{ToA}$}
        \label{fig:ml-toa}
    \end{subfigure}
    \begin{subfigure}{0.3\textwidth}
        \centering
        \begin{tikzpicture}
            \begin{axis}[
                width=0.6\columnwidth,
                height=0.6\columnwidth,
                scale only axis,
                enlargelimits=false,
                axis on top,
                xlabel = {Coordinate $\hat x_1 ~ [\mathrm{m}]$},
                ylabel = {Coordinate $\hat x_2 ~ [\mathrm{m}]$},
                ylabel shift = -8 pt,
                xlabel shift = -4 pt,
                xtick={-15, -10, -5, 0, 5}
            ]
                \addplot graphics[xmin=-16.84,xmax=6.60,ymin=-18.53,ymax=2.96] {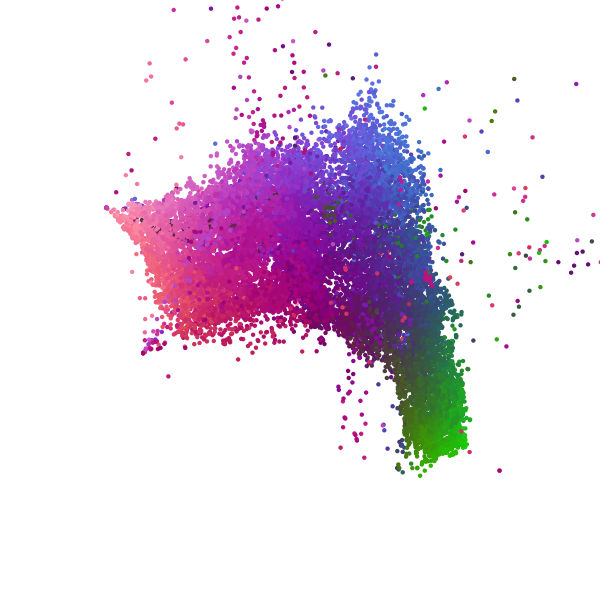};
            \end{axis}
        \end{tikzpicture}
        \vspace{-0.1cm}
        \caption{AoA / Triangulation: $\mathcal L^{(l)}_\mathrm{AoA}$}
        \label{fig:ml-aoa}
    \end{subfigure}
    \begin{subfigure}{0.3\textwidth}
        \centering
        \begin{tikzpicture}
            \begin{axis}[
                width=0.6\columnwidth,
                height=0.6\columnwidth,
                scale only axis,
                enlargelimits=false,
                axis on top,
                xlabel = {Coordinate $\hat x_1 ~ [\mathrm{m}]$},
                ylabel = {Coordinate $\hat x_2 ~ [\mathrm{m}]$},
                ylabel shift = -8 pt,
                xlabel shift = -4 pt,
                xtick={-15, -10, -5, 0, 5}
            ]
                \addplot graphics[xmin=-16.84,xmax=6.60,ymin=-18.53,ymax=2.96] {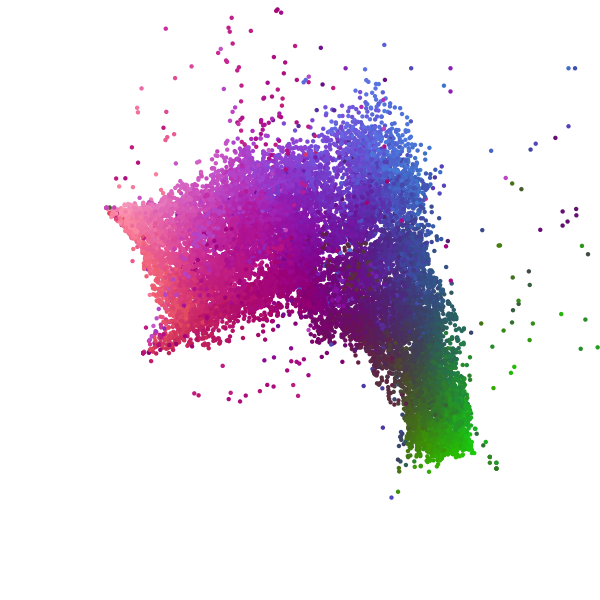};
            \end{axis}
        \end{tikzpicture}
        \vspace{-0.1cm}
        \caption{AoA/ToA Joint Optimization: : $\mathcal L^{(l)}_\mathrm{AoA/ToA}$}
        \label{fig:ml-joint}
    \end{subfigure}
    \caption{Position estimates for the considered classical source localization techniques}
    \label{fig:classical-all}
\end{figure*}

\subsection{Time of Arrival Estimation and Multilateration}
\label{sec:classical-toa}
The two steps involved in \ac{ToA}-based multilateration are to first estimate the \acp{ToA} for all antenna arrays and to then use that information to determine the transmitter location.
Various approaches and algorithms have been proposed for each of these steps.
In our case, we use a root-\ac{MUSIC} super-resolution algorithm for \ac{ToA} estimation, and a \ac{ML}-based approach for the second step.

\subsubsection{Time of Arrival Estimation}
Even though, in theory, every single antenna has its own \ac{ToA}, we assume that the times of arrival within one antenna array are similar enough due to the small antenna spacing and just compute one delay per antenna array.
We proceed as described in \cite{li2004super}: To reduce the computational load, we do not compute the subcarrier autocorrelation matrix $\mathbf R$ over all $N_\mathrm{sub} = 1024$ subcarriers, but split the subcarriers into $U = 4$ segments of length $V = 256$, such that the modified \ac{CSI} tensor for one datapoint can be written as $\mathbf { \tilde H }^{(l)} \in \mathbb C^{B \times M_\mathrm{row} \times M_\mathrm{col} \times U \times V}$.
The autocorrelation matrix $\mathbf R^{(l, b)} \in \mathbb C^{V \times V}$ for antenna array $b$ and time index $l$ is then estimated as
\begin{align*}
    \mathbf {\hat R}^{(l, b)} = &\frac{1}{M_\mathrm{row} \cdot M_\mathrm{col} \cdot U} \cdot \\ &\sum_{m_\mathrm{r} = 1}^{M_\mathrm{row}} \sum_{m_\mathrm{c} = 1}^{M_\mathrm{col}} \sum_{u = 1}^U \mathbf { \tilde H }^{(l)}_{b, m_\mathrm{r}, m_\mathrm{c}, u} \left(\mathbf {\tilde H}^{(l)}_{b, m_\mathrm{r}, m_\mathrm{c}, u}\right)^\mathrm{H},
\end{align*}
where $\mathbf {\tilde H}^{(l)}_{b, m_\mathrm{r}, m_\mathrm{c}, u} \in \mathbb C^V$.
We found that the accuracy of \ac{ToA} estimates improves when applying the \ac{FBCM} method as mentioned in \cite{li2004super}.
We therefore compute the \ac{FBCM} corresponding to $\mathbf {\hat R}^{(l, b)}$ as
\[
    \mathbf {\hat R}_\mathrm{FBCM}^{(l, b)} = \frac{1}{2} \left(\mathbf {\hat R}^{(l, b)} + \mathbf J \left(\mathbf {\hat R}^{(l, b)}\right)^* \mathbf J \right),
\]
where $(\cdot)^*$ symbolizes the complex conjugate, and $\mathbf J$ is the $V \times V$ exchange matrix, where all entries are equal to zero except for those on the anti-diagonal, which are equal to one.
We determine the number of sources (i.e., number of multipath components), which is a required input for \ac{MUSIC}, using a Rissanen \ac{MDL} criterion as in \cite{wax1985detection}.
We then proceed to find the powers and \acp{ToA} of all sources from $\mathbf {\hat R}_\mathrm{FBCM}^{(l, b)}$ using the root-\ac{MUSIC} algorithm.
Out of a subset of a few strongest sources (highest receive powers), we pick the earliest one, assuming that it corresponds to the \ac{LoS} propagation path, if one exists, or to the shortest propagation path otherwise.

The result of the procedure are \ac{ToA} estimates $\hat \tau^{(l, b)}$, with $B = 4$ estimates (one for each antenna array) for every time instance $l$.
It is important to stress that while the \acp{ToA} can be estimated, the \ac{ToT} $\tau_\mathrm{TX}^{(l)}$ is unknown, as receivers and transmitter are not synchronized.

\subsubsection{Maximum Likelihood Multilateration}
Now that \ac{ToA} estimates $\hat \tau^{(l, b)}$ are available, we can define a function that determines the likelihood that the transmitter is located at any particular position $\mathbf x \in \mathbb R^3$.
As an abbreviation, we introduce
\[
    d^{(b)}(\mathbf x) = \left\lVert \mathbf x - \mathbf z^{(b)} \right\rVert_2,
\]
which is the Euclidean distance between the transmitter location $\mathbf x$ and antenna array $b$ located at $\mathbf z^{(b)}$, and let $c_0$ denote the speed of light.
These are two ways to deal with the unknown \ac{ToT} $\tau_\mathrm{TX}^{(l)}$:

\paragraph{Treat $\tau_\mathrm{TX}^{(l)}$ as a nuisance parameter}
Inspired by \cite{henninger2022probabilistic}, we model the \ac{ToA} estimation error as a normally distributed random variable and derive the likelihood function as
\begin{equation}
    \begin{split}
        & \mathcal L_\mathrm{ToA}^{(l)}\left(\mathbf x, \tau_\mathrm{TX}\right) \\=& \prod_{b = 1}^{B} \frac{1}{\sqrt{2 \pi} \sigma^{(l, b)}} \cdot \mathrm{exp} \left( -\frac{1}{2} \left( \frac{\frac{d^{(b)}(\mathbf x)}{c_0} - \hat \tau^{(l, b)} + \tau_\mathrm{TX}}{\sigma^{(l, b)}} \right)^2 \right).
    \end{split}
    \label{eq:toalikelihood}
\end{equation}
Eq. (\ref{eq:toalikelihood}) computes the likelihood that the transmitter is located at $\mathbf x \in \mathbb R^3$ for an assumed \ac{ToT} of $\tau_\mathrm{TX}$.
The variance parameter $\sigma^{(l, b)}$ is determined based on a heuristic that takes into account the observed \ac{RMS} delay spread at array $b$ for datapoint $l$.

\paragraph{Rely on time differences}
Absolute \acp{ToA} $\hat \tau^{(l, b)} - \tau_\mathrm{TX}^{(l)}$ are influenced by the \ac{ToT}.
However, the difference in \ac{ToA}, usually referred to as the \ac{TDoA}, between arrays, is independent of the \ac{ToT}.
Again under the assumption of a normally distributed \ac{ToA} estimation error, we define a \ac{TDoA}-based likelihood function
\begin{equation}
    \begin{split}
        & \mathcal L_\mathrm{TDoA}^{(l)}\left(\mathbf x\right) = \prod_{b_1 = 1}^{B - 1} \prod_{b_2 = b_1 + 1}^{B} \frac{1}{\sqrt{2 \pi} \sigma^{(l, b_1, b_2)}} \\& \cdot \mathrm{exp} \left( -\frac{1}{2} \left( \frac{\frac{d^{(b_2)}(\mathbf x) - d^{(b_1)}(\mathbf x)}{c_0} - (\hat \tau^{(l, b_2)} - \hat \tau^{(l, b_1)})}{\sigma^{(l, b_1, b_2)}} \right)^2 \right)
    \end{split}
    \label{eq:tdoalikelihood}
\end{equation}

where $\sigma^{(l, b_1, b_2)}$ in Eq. (\ref{eq:tdoalikelihood}) is again determined using a heuristic based on the observed \ac{RMS} delay spread of arrays $b_1$ and $b_2$ for datapoint $l$.

After defining $\mathcal L_\mathrm{ToA}^{(l)}$ and $\mathcal L_\mathrm{TDoA}^{(l)}$, we estimate the transmitter location by performing \ac{ML} estimation.
We found that optimizing $\mathcal L_\mathrm{ToA}^{(l)}$ over both $\mathbf x$ and $\tau_\mathrm{TX}$ provides better performance than optimizing $\mathcal L_\mathrm{TDoA}^{(l)}$ over $\mathbf x$:
\[
    \left(\hat {\mathbf x}^{(l)}, \hat \tau_\mathrm{TX}^{(l)}\right) = \argmax_{(\mathbf x, \tau_\mathrm{TX})} \mathcal L_\mathrm{ToA}^{(l)}(\mathbf x, \tau_\mathrm{TX})
\]

We use SciPy's implementation of the \ac{BFGS} algorithm for \ac{ML} optimization.
The \ac{TDoA}-based likelihood function $\mathcal L_\mathrm{TDoA}^{(l)}$ proved useful for initialization:
We initialize the \ac{BFGS} algorithm with a position determined in a previous step by evaluating $\mathcal L_\mathrm{TDoA}^{(l)}$ on a grid of candidate positions $\mathbf x$ and picking the most likely result.
A more computationally efficient procedure is described in \cite{henninger2022probabilistic}.

\subsubsection{Results}
The colorized location estimates $\mathbf { \hat x }^{(l)}$ obtained through multilateration are shown in Fig. \ref{fig:ml-toa}.
Some overall features of the environment are preserved, but the L-shape is heavily distorted and the estimates appear to be very noisy.
Refer to Section \ref{sec:comparison} for performance metrics and a comparison to the other source localization techniques.

\subsection{Angle of Arrival Estimation and Triangulation}
\label{sec:classical-aoa}
Triangulation determines the transmitter position using the \acp{AoA} of the received \ac{LoS} components at all antenna arrays (assuming a \ac{LoS} channel even exists).
In the following, we first extract the \ac{LoS} component from the channel tensor $\mathbf H^{(l)}$, we then estimate all \acp{AoA} using \ac{MUSIC} and finally determine the transmitter location, again using \ac{ML} optimization.
We restrict ourselves to using the azimuth \ac{AoA}, as the elevation \ac{AoA} estimate would not be very accurate (only two rows in the antenna arrays) and also would not contribute much due to the overall geometry of the setup.

\subsubsection{Line of Sight Component Extraction}
We found that the \ac{ToA} estimates derived for multilateration may be used to also improve \ac{AoA} estimation.
Since estimated \acp{ToA} $\hat \tau^{(l, b)}$ should ideally indicate the time of arrival of the \ac{LoS} component, we can use them to isolate said component from all the other multipath components in the \ac{CIR}:
\begin{equation}
    H'^{(l)}_{b, m_\mathrm{r}, m_\mathrm{c}} = \sum_{n = 1}^{N_\mathrm{sub}} \exp \left( \mathrm j 2 \pi \hat \tau^{(l, b)} \frac{n - \nicefrac{N_\mathrm{sub}}{2} - 1}{N_\mathrm{sub}} \right) H^{(l)}_{b, m_\mathrm{r}, m_\mathrm{c}, n}
    \label{eq:extract-toa}
\end{equation}
In essence, Eq. (\ref{eq:extract-toa}) is derived from the inverse Fourier transform and evaluates the reconstructed continuous-time \ac{CIR} at time $\hat \tau^{(l, b)}$.
The results are collected in vectors $\mathbf H'^{(l)}_{b, m_\mathrm{r}} \in \mathbb C^{M_\mathrm{row}}$, which contain the \ac{LoS} component channel coefficient (if a \ac{LoS} exists) for all antennas in row $m_\mathrm{r}$.

\subsubsection{Angle of Arrival Estimation}
Next, for each datapoint $l$ and antenna array $b$, the array correlation matrix $\mathbf A^{(l, b)}$ is estimated over all rows of the \ac{UPA}.
That is, $\mathbf A^{(l, b)}$ may be used to estimate the azimuth \ac{AoA}, but contains no information about the elevation \ac{AoA}:
\[
    \mathbf { \hat A}^{(l, b)} = \sum_{m_\mathrm{r} = 1}^{M_\mathrm{r}} \mathbf H'^{(l)}_{b, m_\mathrm{r}} \left( \mathbf H'^{(l)}_{b, m_\mathrm{r}} \right)^\mathrm{H} \in \mathbb C^{M_\mathrm{c} \times M_\mathrm{c}}
\]
Under the assumption of a single source, we apply the root-\ac{MUSIC} algorithm to $\mathbf { \hat A}^{(l, b)}$ (without \ac{FBCM} computation) to find azimuth \ac{AoA} estimates $\hat \alpha^{(l, b)}$ from $\mathbf { \hat A}^{(l, b)}$ for each datapoint and each antenna array.
An azimuth \ac{AoA} of $\hat \alpha^{(l, b)} = 0^\circ$ indicates that the signal arrives directly from the front.

\subsubsection{Maximum Likelihood Trilateration}
We derive an \ac{AoA}-based likelihood function under the assumption that angle estimation errors are \emph{von Mises}-distributed.
We denote by $\angle_\mathrm{az}(\mathbf x - \mathbf z^{(b)}, \mathbf n^{(b)})$ the azimuth angle between $\mathbf x - \mathbf z^{(b)}$ (the transmitter position relative to antenna array $b$) and $\mathbf n^{(b)}$ (the normal vector of antenna array $b$), i.e., the ideal \ac{AoA} that corresponds to transmitter position $\mathbf x$ for antenna array $b$.
With $I_0$ denoting the modified Bessel function of the first kind of order 0, the \ac{AoA} likelihood function is
\begin{equation}
    \mathcal L_\mathrm{AoA}^{(l)}(\mathbf x) = \prod_{b = 1}^B \frac{\exp \left( \kappa^{(l, b)} \cos \left( \angle_\mathrm{az}(\mathbf x - \mathbf z^{(b)}, \mathbf n^{(b)}) - \hat \alpha^{(l, b)} \right) \right)}{2 \pi I_0(\kappa^{(l, b)})},
    \label{eq:aoalikelihood}
\end{equation}
where $\kappa^{(l, b)}$ is a concentration parameter, which is also heuristically derived from the observed delay spread at array $b$ for datapoint $l$.
As before, the transmitter position estimate is obtained by \ac{ML} optimization using the \ac{BFGS} algorithm:
\[
    \mathbf { \hat x }^{(l)} = \argmax_{\mathbf x} \mathcal L^{(l)}_\mathrm{AoA}(\mathbf x)
\]

\subsubsection{Results}
The colorized location estimates $\mathbf { \hat x }^{(l)}$ obtained through triangulation are shown in Fig. \ref{fig:ml-aoa}.
The L-shaped measurement environment is clearly visible, but the position estimate is clearly wrong for some datapoints, which appear as outliers.
Refer to Section \ref{sec:comparison} for performance metrics and a comparison to the other source localization techniques.

\subsection{Joint Multilateration and Triangulation}
\label{sec:classical-joint}
The \ac{ML} approach to position estimation can easily take into account information from multiple likelihood functions.
\subsubsection{Joint Likelihood}
We define joint likelihood functions
\begin{align*}
    \mathcal L_\mathrm{AoA/TDoA}^{(l)}(\mathbf x) &= \mathcal L_\mathrm{AoA}^{(l)}(\mathbf x) \cdot \mathcal L_\mathrm{TDoA}^{(l)}(\mathbf x) ~ \text{and} \\
    \mathcal L_\mathrm{AoA/ToA}^{(l)}(\mathbf x, \tau_\mathrm{TX}) &= \mathcal L_\mathrm{AoA}^{(l)}(\mathbf x) \cdot \mathcal L_\mathrm{ToA}^{(l)}(\mathbf x, \tau_\mathrm{TX}),
\end{align*}
which take into account both \ac{ToA} and \ac{AoA} estimates.

\subsubsection{Joint Maximum Likelihood Optmization}
We evaluate $\mathcal L_\mathrm{AoA/TDoA}^{(l)}$ on a grid of candidate positions and pick the most likely position to initialize the \ac{BFGS} algorithm that maximizes $\mathcal L_\mathrm{AoA/ToA}^{(l)}$ over $\mathbf x$, with nuisance parameter $\tau_\mathrm{TX}$:
\[
    \left(\hat {\mathbf x}^{(l)}, \hat \tau_\mathrm{TX}^{(l)}\right) = \argmax_{(\mathbf x, \tau_\mathrm{TX})} \mathcal L_\mathrm{AoA/ToA}^{(l)}(\mathbf x, \tau_\mathrm{TX})
\]

\subsubsection{Results}
The colorized location estimates $\mathbf { \hat x }^{(l)}$ obtained through joint triangulation and multilateration are shown in Fig. \ref{fig:ml-joint}.
The L-shape of the environment is preserved, but the overall quality of position estimates appears to be very similar to the purely triangulation-based method (Fig. \ref{fig:ml-aoa}), with somewhat fewer outliers.
A more detailed analysis of performance metrics in Section \ref{sec:comparison} reveals that joint triangulation and multilateration slightly improves performance over just triangulation, as is to be expected.

\section{Channel Charting-based Localization}
\label{sec:cc}

\begin{figure*}
    \centering
    \begin{subfigure}[t]{0.3\textwidth}
        \centering
        \begin{tikzpicture}
            \begin{axis}[
                width=0.6\columnwidth,
                height=0.6\columnwidth,
                scale only axis,
                enlargelimits=true,
                axis on top,
                xlabel = {Coordinate $\hat x_1$},
                ylabel = {Coordinate $\hat x_2$},
                ylabel shift = -8 pt,
                xlabel shift = -4 pt
            ]
                \addplot[thick,blue] graphics[xmin=-397.08453369140625, xmax=288.62841796875, ymin=-285.5689697265625, ymax=460.52691650390625] {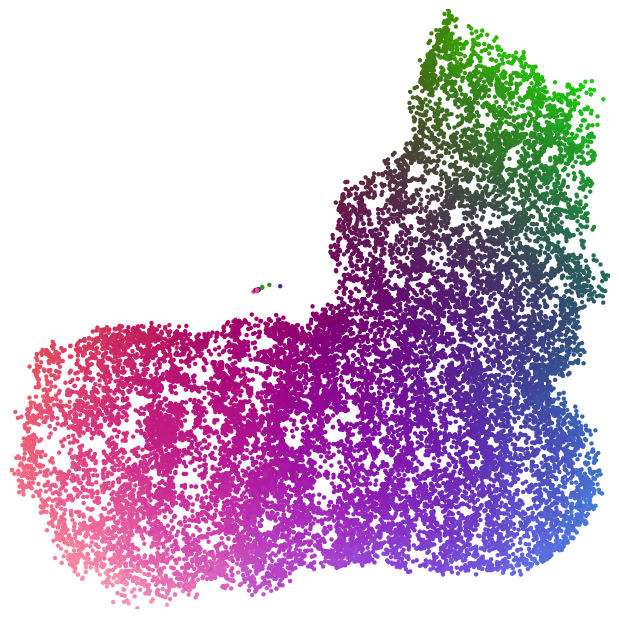};
            \end{axis}
        \end{tikzpicture}
        \vspace{-0.1cm}
        \caption{Channel chart learned by $\mathcal C_\theta$}
        \label{fig:cc}
    \end{subfigure}
    \hspace{0.02\textwidth}
    \begin{subfigure}[t]{0.3\textwidth}
        \centering
        \begin{tikzpicture}
            \begin{axis}[
                width=0.6\columnwidth,
                height=0.6\columnwidth,
                scale only axis,
                enlargelimits=false,
                axis on top,
                xlabel = {Coordinate $\hat x_1 ~ [\mathrm{m}]$},
                ylabel = {Coordinate $\hat x_2 ~ [\mathrm{m}]$},
                ylabel shift = -8 pt,
                xlabel shift = -4 pt
            ]
                \addplot graphics[xmin=-13.028850841408582, xmax=3.6687306000188045, ymin=-15.031928608687338, ymax=-1.0398224119322146] {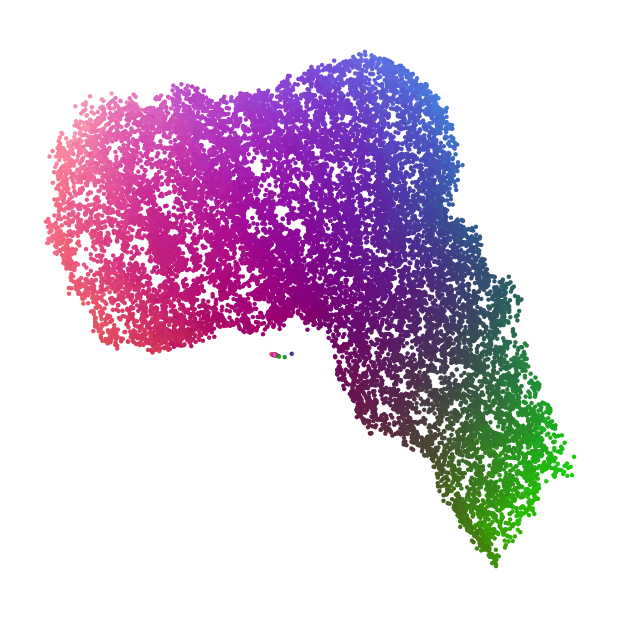};
            \end{axis}
        \end{tikzpicture}
        \vspace{-0.1cm}
        \caption{Channel chart learned by $\mathcal C_\theta$ after optimal affine transform $T_\mathrm{c} \circ \mathcal C_\theta$}
        \label{fig:cc-affine}
    \end{subfigure}
    \hspace{0.02\textwidth}
    \begin{subfigure}[t]{0.3\textwidth}
        \centering
        \begin{tikzpicture}
            \begin{axis}[
                width=0.6\columnwidth,
                height=0.6\columnwidth,
                scale only axis,
                enlargelimits=false,
                axis on top,
                xlabel = {Coordinate $\hat x_1 ~ [\mathrm{m}]$},
                ylabel = {Coordinate $\hat x_2 ~ [\mathrm{m}]$},
                ylabel shift = -8 pt,
                xlabel shift = -4 pt
            ]
                \addplot[thick,blue] graphics[xmin=-13.02411937713623, xmax=2.807442307472229, ymin=-15.318949699401855, ymax=-0.49260711669921875] {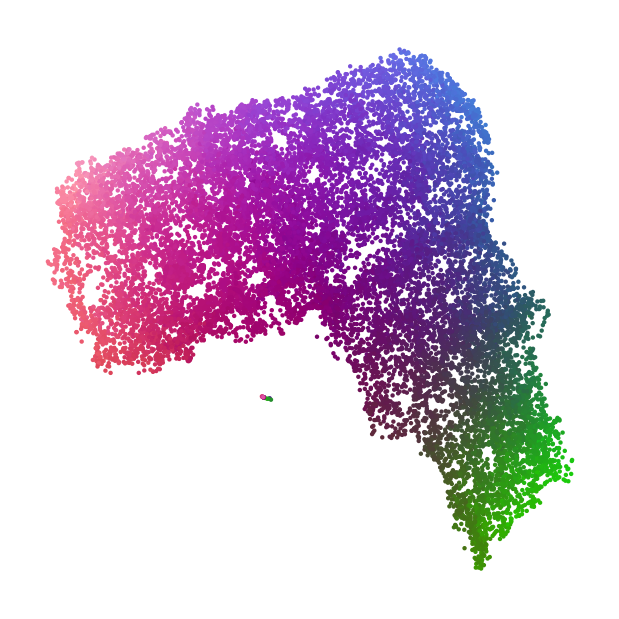};
            \end{axis}
        \end{tikzpicture}
        \vspace{-0.1cm}
        \caption{Augmented channel chart learned by $\mathcal C_{\theta, \mathrm{aug}}$}
        \label{fig:cc-augmented}
    \end{subfigure}
    \caption{Position estimates for Channel Charting-based source localization techniques}
    \label{fig:cc-all}
\end{figure*}

\subsection{Dissimilarity Metric}
We use an approach that we call \emph{dissimilarity metric-based} Channel Charting and that we previously described in \cite{stephan2023angle}.
In the following, we use our \emph{geodesic, fused} dissimilarity metric, which combines information extracted from the \ac{ADP} and from timestamp differences.
Refer to \cite{stephan2023angle} for information on how this metric is computed.
We denote the dissimilarity between the datapoints $l_\mathrm{x}$ and $l_\mathrm{y}$ according to said metric by $d_\mathrm{G-fuse}^{(l_\mathrm{x}, l_\mathrm{y})}$.

\subsection{Siamese Neural Network}
We implement the \ac{FCF} $\mathcal C_\theta: \mathbb C^{B \times M_\mathrm{row} \times M_\mathrm{col} \times N_\mathrm{sub}} \to \mathbb R^2$, which maps \ac{CSI} tensors $\mathbf H^{(l)}$ to channel chart positions $\mathbf x^{(l)} = \mathcal C_\theta\left(\mathbf H^{(l)}\right)$, as a \ac{NN}.
The \ac{NN} is trained as a Siamese neural network, as in \cite{dissim-cc-tutorial}.
The Siamese loss function that we use is inspired by Sammon's loss \cite{sammon_mapping} and is defined in terms of two datapoints with indices $l_\mathrm{x}$ and $l_\mathrm{y}$, whose channel chart coordinates are $\mathbf x = \mathcal C_\theta\left(\mathbf H^{(l_\mathrm{x})}\right)$ and $\mathbf y = \mathcal C_\theta\left(\mathbf H^{(l_\mathrm{y})}\right)$:
\begin{equation}
    \mathcal{L}_\mathrm{siam}^{(l_\mathrm{x}, l_\mathrm{y})}(\mathbf x, \mathbf y) = \frac{\left(d_\mathrm{G-fuse}^{(l_\mathrm{x}, l_\mathrm{y})} - \Vert\mathbf y - \mathbf x\Vert_2\right)^2}{d_\mathrm{G-fuse}^{(l_\mathrm{x}, l_\mathrm{y})} + \beta}
    \label{eq:siameseloss}
\end{equation}
$\mathcal{L}_\mathrm{siam}^{(l_\mathrm{x}, l_\mathrm{y})}$ describes an error measure between the Euclidean distance of predicted channel chart locations $\lVert \mathbf y - \mathbf x \rVert_2$ and the computed dissimilarity $d_\mathrm{G-fuse}^{(l_\mathrm{x}, l_\mathrm{y})}$.
The parameter $\beta$ is a hyperparameter that can tune Eq. (\ref{eq:siameseloss}) to weight either the absolute squared error (for large $\beta$) or the normalized squared error (for small $\beta$) higher.
By evaluating the loss function over a batch of datapoint pairs $(l_\mathrm{x}, l_\mathrm{y})$ and backpropagation of gradients, the Siamese \ac{NN} learns to map high-dimensional \ac{CSI} to channel chart locations such that channel chart distances correspond best to the previously computed dissimilarities in the sense of Eq. (\ref{eq:siameseloss}).

\subsection{Intermediate Results}
By applying the learned forward charting function $\mathcal C_\theta$ to all channel tensors $\mathbf H^{(l)}$ in the dataset $\mathcal S$, we obtain the channel chart shown in Fig. \ref{fig:cc}.
While the channel chart preserves local datapoint neighborhood relationships and also reproduces the global L-shaped topology of the environment, it is arbitrarily scaled, rotated and flipped compared to the ``ground truth'' positions in Fig. \ref{fig:groundtruth-map}.
This intuitively makes sense, since the \ac{FCF} has no way of knowing anything about the absolute scale or orientation of the environment, as it is only trained on dissimilarities.
Inspired by \cite{pihlajasalo2020absolute} and \cite{fraunhofer_cc}, we suggest to use the position estimates obtained through classical techniques (Section \ref{sec:classical}) to estimate an affine transform $T_\mathrm{c}$ from the channel chart's coordinate frame to physical coordinates.

\subsection{Optimal Linear Transform}
\label{sec:cc-affine}
To find $T_\mathrm{c}(\mathbf x) = \hat { \mathbf A } \mathbf x + \hat { \mathbf b }$, we solve the least squares problem
\[
    (\mathbf{\hat A}, \mathbf{\hat b}) = \argmin\limits_{(\mathbf{A}, \mathbf{b})} \sum_{l = 1}^L \lVert\mathbf{A} \hat {\mathbf x }_\mathrm{CC}^{(l)} + \mathbf b - \hat {\mathbf x }^{(l)}_\mathrm{AoA/ToA} \rVert_2^2,
\]
where $\hat {\mathbf x }^{(l)}_\mathrm{CC} = \mathcal C_\theta\left(\mathbf H^{(l)}\right)$ denotes the position estimate obtained through Channel Charting, and $\hat {\mathbf x }^{(l)}_\mathrm{AoA/ToA}$ denotes the position estimate obtained by \ac{ML} optimization of $\mathcal L_\mathrm{AoA/ToA}^{(l)}$.
We write the composition of $T_\mathrm{c}$ and $C_\theta$, i.e., Channel Charting and subsequent application of $T_\mathrm{c}$ as $T_\mathrm{c} \circ C_\theta$.

\subsection{Results}
The channel chart after the optimal affine transform is shown in Fig. \ref{fig:cc-affine}.
Clearly, the chart in Fig. \ref{fig:cc-affine} is identical to the one in Fig. \ref{fig:cc}, except for the affine transform.
The channel chart very much resembles the L-shaped map of ``ground truth'' positions, and orientation and scale now also match the reference.

As an aside, the learned \acp{FCF} $\mathcal C_\theta$ and $\mathcal C_{\theta, \mathrm{aug}}$ were evaluated on $\mathcal S$, i.e., the same dataset that they were trained on.
Since Channel Charting is a self-supervised learning technique, a test set is not strictly needed to get a meaningful performance evaluation.
In some additional experiments, we also evaluated both $\mathcal C_\theta$ and $\mathcal C_{\theta, \mathrm{aug}}$ on a test set of strictly different datapoints that was measured in the same environment and did not observe a significant drop in performance.

\section{Augmented Channel Charting}
\label{sec:augmented}
In Section \ref{sec:cc}, Channel Charting and transforming channel chart coordinates into physical coordinates using classical position estimates were regarded as two separate steps.
This has several disadvantages:
For one, if there is a non-affine distortion in the channel chart coordinates, the affine transform $T_\mathrm{c}$ cannot correct for it, even if the distortion is not present in the classical position estimates.
This effect is faintly visible in Fig. \ref{fig:cc-affine}: The angle between the two sides of the L-shape is too obtuse compared to the ``ground truth'' positions in \ref{fig:groundtruth-map}, even though the classical estimates in Fig. \ref{fig:ml-joint} clearly show a square angle.
For another, the \ac{NN} can easily take into account classical \ac{AoA} and \ac{ToA} estimates during training, without first reducing the information to position estimates.
By factoring in similarity relationships, \ac{ToA} estimates and also \ac{AoA} estimates while training the \ac{FCF}, all available information can be supplied to the \ac{NN} at once.
This is the idea behind what we will refer to as \emph{augmented} Channel Charting, which we will explain in the following.

\subsection{Dissimilarity Matrix Scaling}
Before all sources of information (\ac{AoA} / \ac{ToA} / dissimilarities) can be taken into account in a unified framework, we first need to ensure compatibility.
The dissimilarities $d_\mathrm{G-fuse}^{(l_\mathrm{x}, l_\mathrm{y})}$ are computed based on \ac{ADP} and timestamps.
While $d_\mathrm{G-fuse}^{(l_\mathrm{x}, l_\mathrm{y})}$ should ideally be proportional to the true Euclidean distance $\left\lVert \mathbf x^{(l_\mathrm{y})} - \mathbf x^{(l_\mathrm{x})} \right\rVert_2$ between datapoints $l_\mathrm{x}$ and $l_\mathrm{y}$, there remains an unknown scaling factor $\gamma$, i.e., $d_\mathrm{G-fuse}^{(l_\mathrm{x}, l_\mathrm{y})} \approx \gamma \left\lVert \mathbf x^{(l_\mathrm{y})} - \mathbf x^{(l_\mathrm{x})} \right\rVert_2$.
We estimate $\gamma$ as
\begin{equation}
    \hat \gamma = \frac{1}{L^2} \sum_{l_\mathrm{x} = 1}^L \sum_{l_\mathrm{y} = 1}^L \frac{d_\mathrm{G-fuse}^{(l_\mathrm{x}, l_\mathrm{y})}}{\left\lVert \hat {\mathbf x}_\mathrm{AoA/ToA}^{(l_\mathrm{y})} - \hat {\mathbf x}_\mathrm{AoA/ToA}^{(l_\mathrm{x})} \right\rVert_2},
    \label{eq:scalingfactor}
\end{equation}
where $\hat {\mathbf x}_\mathrm{AoA/ToA}^{(l)}$ is again the classical position estimate obtained by \ac{ML} optimization of $\mathcal L_\mathrm{AoA/ToA}^{(l)}$.
Instead of estimating $\gamma$ over all pairs of datapoints $(l_\mathrm{x}, l_\mathrm{y})$, Eq. (\ref{eq:scalingfactor}) may also be applied to a random subset of datapoint pairs to reduce computational complexity.
We write the dissimilarities scaled by $\gamma$ as
\[
    \tilde d_\mathrm{G-fuse}^{(l_\mathrm{x}, l_\mathrm{y})} = \frac{1}{\gamma} d_\mathrm{G-fuse}^{(l_\mathrm{x}, l_\mathrm{y})} \approx \left\lVert \mathbf x^{(l_\mathrm{y})} - \mathbf x^{(l_\mathrm{x})} \right\rVert_2.
\]

\subsection{Combined Loss Function}
The key element of augmented Channel Charting is the \emph{combined} loss function, which takes into account both a Siamese loss (similar to Eq. (\ref{eq:siameseloss}), but without normalization) and the classical \ac{AoA} / \ac{TDoA}-based likelihood function:
\begin{equation}
    \begin{split}
        \mathcal{L}_\mathrm{comb}^{(l_\mathrm{x}, l_\mathrm{y})}(\mathbf x, \mathbf y) = (1 - \lambda) &\left( \tilde d_\mathrm{G-fuse}^{(l_\mathrm{x}, l_\mathrm{y})} - \lVert \mathbf y - \mathbf x \rVert_2\right)^2 \\- \lambda &\left(\mathcal L^{(l_\mathrm{y})}_\mathrm{AoA/TDoA}(\mathbf y) + \mathcal L^{(l_\mathrm{x})}_\mathrm{AoA/TDoA}(\mathbf x)\right)
    \end{split}
    \label{eq:combinedloss}
\end{equation}
Note that minimizing $\mathcal{L}_\mathrm{comb}^{(l_\mathrm{x}, l_\mathrm{y})}$ simultaneously minimizes the Siamese loss given by the squared error term $( \tilde d_\mathrm{G-fuse}^{(l_\mathrm{x}, l_\mathrm{y})} - \lVert \mathbf y - \mathbf x \rVert_2)^2$ and maximizes the classical likelihood function $\mathcal L_\mathrm{AoA/TDoA}^{l}$ for both datapoints $l_\mathrm{x}$ and $l_\mathrm{y}$.
The weighting of these two objectives is determined by the weighting factor $\lambda$, $0 \leq \lambda \leq 1$, which is a hyperparameter that may be adjusted over the training period.
For example, $\lambda$ may be assigned a high value in the beginning to quickly capture the global geometry, and lower values in subsequent \ac{NN} training steps.

We want to point out that there is room for improvement for the combined loss function, in the sense that Eq. (\ref{eq:combinedloss}) mixes likelihood functions and a squared error term (which may be interpreted as a negative log-likelihood function).
We chose to define the combined loss as given in Eq. (\ref{eq:combinedloss}) as it produced convincing results in our experiments.

\subsection{Results}
The channel chart learned with augmented Channel Charting is shown in Fig. \ref{fig:cc-augmented}.
It is similar to the one in Fig. \ref{fig:cc-affine}, but the edges of the shape are now straighter and there is a right angle between the two sides of the L-shape, just like in the ``ground truth'' map.
Overall, the channel chart looks very similar to the reference data.
The performance metrics in Section \ref{sec:comparison} show that augmented Channel Charting outperforms all the other source localization techniques.

\section{Results, Comparison and Conclusion}
\label{sec:comparison}
\begin{table*}
    \centering
    \caption{Performance comparison of all classical (maximum likelihood-based) and Channel Charting-based methods}
    \vspace{-0.1cm}
    \begin{tabular}{r c | c c c c c c}
        & \textbf{Loss} & \textbf{MAE $\downarrow$} & \textbf{DRMS $\downarrow$} & \textbf{CEP $\downarrow$} & \textbf{R95 $\downarrow$} & \textbf{KS $\downarrow$} & \textbf{CT/TW $\uparrow$} \\ \hline
        ML: ToA & $\mathcal L_\mathrm{ToA}^{(l)}$ & $2.314\,\mathrm m$ & $3.090\,\mathrm m$ & $1.715\,\mathrm m$ & $6.298\,\mathrm m$ & $0.980$ & $0.871/0.933$ \\
        ML: AoA & $\mathcal L_\mathrm{AoA}^{(l)}$ & $0.909\,\mathrm m$ & $1.643\,\mathrm m$ & $0.574\,\mathrm m$ & $2.563\,\mathrm m$ & $0.236$ & $0.932/0.936$ \\
        ML: AoA / ToA & $\mathcal L_\mathrm{AoA/ToA}^{(l)}$ & $0.676\,\mathrm m$ & $1.228\,\mathrm m$ & $0.462\,\mathrm m$ & $1.763\,\mathrm m$ & $0.214$ & $0.965/0.970$ \\ \hline
        CC: $T_\mathrm{c} \circ C_\theta$ & $\mathcal L_\mathrm{siam}^{(l)}$ & $0.490\,\mathrm m$ & $0.584\,\mathrm m$ & $0.441\,\mathrm m$ & $1.026\,\mathrm m$ & $0.071$ & $\mathbf{0.996/0.996}$ \\
        CC: $C_{\theta, \mathrm{aug}}$ & $\mathcal L_\mathrm{aug}^{(l)}$ & $\mathbf{0.401\,m}$ & $\mathbf{0.483\,m}$ & $\mathbf{0.369\,m}$ & $\mathbf{0.789\,m}$ & $\mathbf{0.070}$ & $0.995/0.995$
    \end{tabular}
    \label{tab:performance}
    \vspace{-0.3cm}
\end{table*}

\begin{figure}
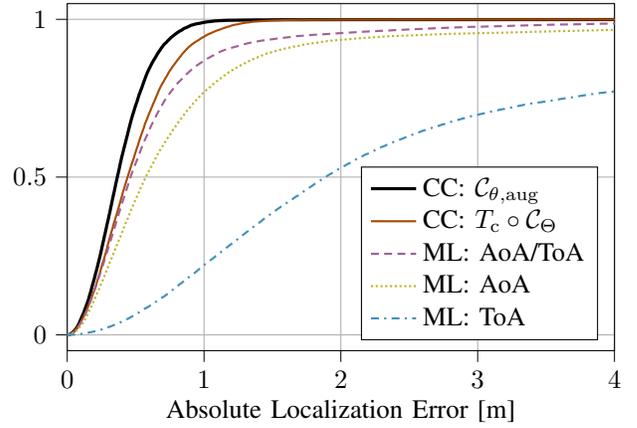

    \centering
    \include{fig/cdf}
    \vspace{-0.4cm}
    \caption{Empirical cumulative distribution functions of absolute localization errors for all classical (maximum likelihood-based) and Channel Charting-based methods}
    \label{fig:cdf}
\end{figure}

The top view maps for different localization techniques are shown in Fig. \ref{fig:classical-all} and Fig. \ref{fig:cc-all}.
Clearly, Fig. \ref{fig:cc-augmented} approximates the ``ground truth'' positions in Fig. \ref{fig:groundtruth-map} best and also exhibits the smallest number of outlier estimates.
To compare estimation techniques more objectively beyond this visual intuition, we introduce a set of formal performance criteria, which are commonly used in literature to quantify localization performance.
We write the ``ground truth'' positions as $\mathbf x^{(l)}$ and the position estimates as $\hat {\mathbf x}^{(l)}$ and compute the metrics over the complete dataset $\mathcal S$.
First of all, the \ac{MAE} and \ac{DRMS} are defined as
\[
    \begin{split}
        \mathrm{MAE} &= \frac{1}{L} \sum_{l=1}^L \lVert \mathbf x^{(l)} - \hat {\mathbf x}^{(l)} \rVert_2 ~ \text{and} \\
        \mathrm{DRMS} &= \sqrt{\frac{1}{L} \sum_{l=1}^L \lVert \mathbf x^{(l)} - \hat {\mathbf x}^{(l)} \rVert_2^2}.
    \end{split}
\]
Furtheremore, we define the \ac{CEP} and the \ac{R95} as the median and 95\textsuperscript{th} percentile of the empirical distribution of the localization error $\left\lVert \mathbf x^{(l)} - \hat {\mathbf x}^{(l)} \right\rVert_2$, and also consider the \ac{eCDF} of this error.
Finally, we borrow the well-known dimensionality reduction performance metrics \ac{CT} and \ac{TW} (level of preservation of local geometry), and \ac{KS} (level of preservation of global geometry) from Channel Charting literature, as defined, e.g., in \cite{stephan2023angle} or \cite{fraunhofer_cc}.
We abbreviate the classical \ac{ML}-based approach solely relying on \ac{ToA} information (as in \ref{sec:classical-toa}) as ``ML: ToA'', triangulation, which solely uses \ac{AoA} information (as in \ref{sec:classical-aoa}) as ``ML: AoA'' and the classical approach with the joint likelihood function (as in \ref{sec:classical-joint}) as ``ML: AoA/ToA''.
We abbreviate the Channel Charting-based solution that applies the optimal affine transform (as in \ref{sec:cc-affine}) as ``CC: $T_\mathrm{c} \circ C_\theta$'' and augmented Channel Charting as ``CC: $\mathcal C_{\theta, \mathrm{aug}}$.

The \acp{eCDF} in Fig. \ref{fig:cdf} show, unsurprisingly, that joint (\ac{AoA}/\ac{ToA}) classical \ac{ML} optimization outperforms pure triangulation or multilateration.
Pure \ac{AoA}-based triangulation is much better than pure \ac{ToA}-based multilateration, which is expected given the relatively low signal bandwidth and the small geometric scale of the measurement environment.
Most importantly, though, the \acp{eCDF} in Fig. \ref{fig:cdf} also show that the two Channel Charting-based methods clearly outperform all classical techniques.
The same conclusion can be arrived at by comparing the performance metrics in Tab. \ref{tab:performance}:
The two Channel Charting-based techniques outperform the classical methods with respect to all metrics.
Augmented Channel Charting outperforms regular Channel Charting with subsequent affine transform with respect to all metrics except for \ac{CT} and \ac{TW}.
However, the difference in \ac{CT} / \ac{TW}, and hence preservation of local geometry, is negligible compared to the significant localization performance gains of augmented Channel Charting over regular Channel Charting.

This leads us to the following conclusion:
Not only can Channel Charting be enhanced by taking into account information from classical localization techniques, but Channel Charting can also be used to improve overall localization accuracy.
One might argue that there exist better classical localization techniques than just the ones presented here (e.g., ones that also take into account received powers, elevation angles or phase differences of arrival between different arrays).
If that is the case, the performance gap between Channel Charting and these classical techniques may shrink.
However, with the framework of augmented Channel Charting, the additional information derived through classical techniques may be taken into account while training the \ac{FCF}.
With a suitable \ac{NN} architecture, suitable features and sufficient training data, the universal approximation theorem should thus ensure that the learned \ac{FCF} is always at least as good as the classical solution.

\section{Outlook}
We not only demonstrated that Channel Charting can outperform classical source localization techniques in terms of localization performance, but also showed that Channel Charting may be improved by taking into account information derived through classical \ac{AoA} or \ac{ToA}-based approaches.
To this end, the augmented Channel Charting framework was put forward, which, in the future, may also take into account other sources of side information available to the system.
While we investigated Channel Charting as a source localization technique in this paper, we want to stress that the improvements in localization performance through augmented Channel Charting may also benefit other applications of Channel Charting, like handover prediction or pilot assignment.

\bibliographystyle{IEEEtran}
\bibliography{IEEEabrv,references}

\begin{thebibliography}{10}
\providecommand{\url}[1]{#1}
\csname url@samestyle\endcsname
\providecommand{\newblock}{\relax}
\providecommand{\bibinfo}[2]{#2}
\providecommand{\BIBentrySTDinterwordspacing}{\spaceskip=0pt\relax}
\providecommand{\BIBentryALTinterwordstretchfactor}{4}
\providecommand{\BIBentryALTinterwordspacing}{\spaceskip=\fontdimen2\font plus
\BIBentryALTinterwordstretchfactor\fontdimen3\font minus
  \fontdimen4\font\relax}
\providecommand{\BIBforeignlanguage}[2]{{%
\expandafter\ifx\csname l@#1\endcsname\relax
\typeout{** WARNING: IEEEtran.bst: No hyphenation pattern has been}%
\typeout{** loaded for the language `#1'. Using the pattern for}%
\typeout{** the default language instead.}%
\else
\language=\csname l@#1\endcsname
\fi
#2}}
\providecommand{\BIBdecl}{\relax}
\BIBdecl

\bibitem{studer_cc}
C.~Studer, S.~Medjkouh, E.~G{\"{o}}n{\"{u}}ltas, T.~Goldstein, and
  O.~Tirkkonen, ``{Channel Charting: Locating Users within the Radio
  Environment using Channel State Information},'' \emph{CoRR}, vol.
  abs/1807.05247, 2018.

\bibitem{taner2023channel}
S.~Taner, V.~Palhares, and C.~Studer, ``{Channel Charting in Real-World
  Coordinates},'' \emph{arXiv preprint arXiv:2308.14498}, 2023.

\bibitem{pihlajasalo2020absolute}
J.~Pihlajasalo, M.~Koivisto, J.~Talvitie, S.~Ali-L{\"o}ytty, and M.~Valkama,
  ``Absolute positioning with unsupervised multipoint channel charting for 5g
  networks,'' in \emph{2020 IEEE 92nd Vehicular Technology Conference
  (VTC2020-Fall)}.\hskip 1em plus 0.5em minus 0.4em\relax IEEE, 2020, pp. 1--5.

\bibitem{stephan2023angle}
P.~Stephan, F.~Euchner, and S.~ten Brink, ``{Angle-Delay Profile-Based and
  Timestamp-Aided Dissimilarity Metrics for Channel Charting},'' \emph{arXiv
  preprint arXiv:2308.09539}, 2023.

\bibitem{dichasus2021}
F.~Euchner, M.~Gauger, S.~D\"orner, and S.~ten Brink, ``{A Distributed Massive
  MIMO Channel Sounder for "Big CSI Data"-driven Machine Learning},'' in
  \emph{WSA 2021; 25th International ITG Workshop on Smart Antennas}, 2021.

\bibitem{dataset-dichasus-cf0x}
\BIBentryALTinterwordspacing
F.~Euchner and M.~Gauger, ``{CSI Dataset dichasus-cf0x: Distributed Antenna
  Setup in Industrial Environment, Day 1},'' 2022. [Online]. Available:
  \url{https://doi.org/doi:10.18419/darus-2854}
\BIBentrySTDinterwordspacing

\bibitem{li2004super}
X.~Li and K.~Pahlavan, ``{Super-resolution TOA estimation with diversity for
  indoor geolocation},'' \emph{IEEE transactions on wireless communications},
  vol.~3, no.~1, pp. 224--234, 2004.

\bibitem{wax1985detection}
M.~Wax and T.~Kailath, ``Detection of signals by information theoretic
  criteria,'' \emph{IEEE Transactions on acoustics, speech, and signal
  processing}, vol.~33, no.~2, pp. 387--392, 1985.

\bibitem{henninger2022probabilistic}
M.~Henninger, T.~E. Abrudan, S.~Mandelli, M.~Arnold, S.~Saur, V.-M. Kolmonen,
  S.~Klein, T.~Schlitter, and S.~Ten~Brink, ``{Probabilistic 5G Indoor
  Positioning Proof of Concept with Outlier Rejection},'' in \emph{2022 Joint
  European Conference on Networks and Communications \& 6G Summit (EuCNC/6G
  Summit)}.\hskip 1em plus 0.5em minus 0.4em\relax IEEE, 2022, pp. 249--254.

\bibitem{dissim-cc-tutorial}
\BIBentryALTinterwordspacing
F.~Euchner, C.~Eun~Lee, and P.~Stephan. (2023) {Dissimilarity Metric-Based
  Channel Charting}. [Online]. Available:
  \url{https://dichasus.inue.uni-stuttgart.de/tutorials/tutorial/dissimilarity-metric-channelcharting/}
\BIBentrySTDinterwordspacing

\bibitem{sammon_mapping}
J.~Sammon, ``{A Nonlinear Mapping for Data Structure Analysis},'' \emph{IEEE
  Transactions on Computers}, vol. C-18, no.~5, pp. 401--409, 1969.

\bibitem{fraunhofer_cc}
M.~Stahlke, G.~Yammine, T.~Feigl, B.~M. Eskofier, and C.~Mutschler, ``{Indoor
  Localization with Robust Global Channel Charting: A Time-Distance-Based
  Approach},'' \emph{IEEE Transactions on Machine Learning in Communications
  and Networking}, pp. 1--1, 2023.

\end{thebibliography}

\end{document}